\documentclass[aip,rsi,reprint,amsmath,amssymb]{revtex4-1}

\usepackage{graphicx}
\usepackage{dcolumn}
\usepackage{bm}
\usepackage{natbib}

\begin{document}

\title{Wide-range wavevector selectivity of magnon gases in Brillouin light scattering spectroscopy}

\author{C.~W. Sandweg}
\email{sandweg@physik.uni-kl.de}
\author{M.~B. Jungfleisch}
\author{V.~I. Vasyuchka}
\author{A.~A. Serga}
\author{P. Clausen}
\author{H. Schultheiss}
\author{B. Hillebrands}
\affiliation{%
Fachbereich Physik and Forschungszentrum OPTIMAS,
Technische Universit\"at Kaiserslautern, 67663 Kaiserslautern, Germany}%
\author{A. Kreisel}
\author{P. Kopietz}
\affiliation{%
Institut f\"ur Theoretische Physik, Universit\"at Frankfurt, 
60438 Frankfurt, Germany
}%
\date{\today}

\begin{abstract}
Brillouin light scattering spectroscopy is a powerful technique for the study of fast magnetization dynamics with both
frequency- and wavevector resolution. Here, we report on a distinct improvement of this spectroscopic technique towards
two-dimensional wide-range wavevector selectivity in a backward scattering geometry. Spin-wave wavevectors 
oriented perpendicular to the bias magnetic field are investigated by tilting the sample within the magnet gap. Wavevectors which 
are oriented parallel to the applied magnetic field are analyzed by turning the entire setup, including the
magnet system. The setup features a wide selectivity of wavevectors up to $2.04\cdot 10^{5}$~rad/cm for both
orientations, and allows selecting and measuring wavevectors of dipole- and exchange-dominated spin waves of any orientation to the magnetization simultaneously.
\end{abstract}

\maketitle

\section{Introduction}

In the last two decades Brillouin light scattering (BLS) spectroscopy became an important optical technique for the
investigation of magnetization dynamics in thin films and in confined elements \cite{Demidov, Schultheiss}. This method allows the determination of the frequency spectrum of the eigen excitations of different magnetic structures \cite{Sandweg}, gives information about the frequencies and wavevectors of spin waves in magnonic crystals \cite{Demo-book}, and is an important tool for investigating the behavior of magnon gases as well as Bose-Einstein condensates (BEC)
of magnons \cite{LSS-book, Dem06, Demidov2008}.

It is well-known that the inelastic Brillouin light scattering process can be interpreted in a classical way as resulting from the reflection of probing light
from a moving Bragg-grating formed by spin waves due to magneto-optical coupling\cite{Dem01}. In such a case two spectral peaks appear, each on one side of the frequency of the probing light, due to the Doppler effect. The frequency shift of these peaks is equal
to the frequency of the spin wave, and their intensity is proportional to the density of magnons. Both characteristics are
detectable by a high-contrast tandem Fabry-Perot interferometer\cite{LSS-book2}.

\begin{figure}
\begin{center}
\scalebox{1}{\includegraphics[width=8.5 cm, clip]{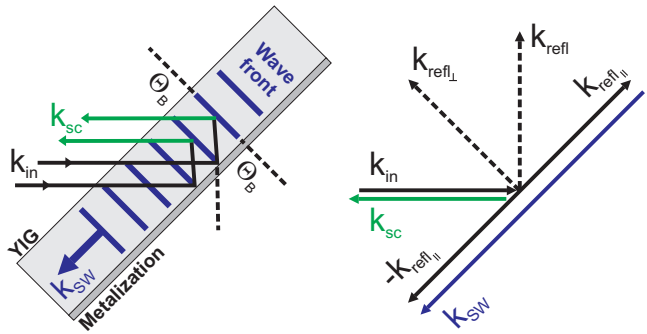}}
\end{center}
\caption{\label{Bragg} (Color online) Bragg grating created by spin waves due to magneto-optical coupling.
The incoming light is reflected at the metallized back-side of the YIG film and then inelastically scatters from the Bragg grating.}
\end{figure}

The direction of propagation of the inelastically scattered light provides information on the wavevector of the spin
wave. The transfer of momentum which happens during the scattering process occurs due to diffraction from the grating as shown in
Fig.~\ref{Bragg}. Consequently, the maximum transfer of momentum can be observed when the Bragg-condition
\begin{equation}\label{eq1}
k_\mathrm{SW} = 2k_\mathrm{in}\sin(\Theta_{B})
\end{equation}
is satisfied. Here, $k_\mathrm{SW}$ is the wavenumber of the spin wave propagating in the plane of the magnetic film, $k_\mathrm{in}$ is the wavenumber of the probing light, and $\Theta_{B}$ is the angle formed by the two wave vectors. The access to the wave numbers permits the full spectral characterization of spin-wave systems. In particular, the detection of exchange-dominated spin waves with large wavenumbers in the range of $1\cdot
10^{5}$~rad/cm is crucial for the investigation of parametrically excited magnons, thermalization processes in magnon
gases, and dynamics of BEC of magnons\cite{Demidov2008}. 

In the context of BLS spectroscopy of magnetic phenomena two different measurement geometries are used. Firstly, in
the forward-scattering geometry, the focused probing light passes through an optically transparent ferrimagnetic sample like an yttrium iron garnet (YIG)
film, is collected by an objective lens, and finally directed to the interferometer. The wavevector selectivity can
be realized in this case using a movable diaphragm between the objective with a high numerical aperture and the interferometer. Depending
on the size, the shape, and the position of the hole of the diaphragm, some components of the scattered light and
consequently, certain in-plane wavevectors can be detected, while others are blocked \cite{Wilber, Demidov2008, Kabos, Kabos2}. This
technique allows easy realization without any change of the existing setup, but only wavevectors up to $4.2\cdot
10^{4}$~rad/cm can be detected \cite{Demidov, Demidov2008, Cesar, Neumann}.

The second technique is backward-scattering, wherein the probing beam is focused at the sample, and the backscattered
light is collected by the same objective. This geometry is commonly used for the study of optically opaque samples, e.g metal films \cite{Eremenko, Gubbiotti}, or YIG samples with a metalized backside\cite{Dem06, Serga2007, Serga}.
In the latter case metalization is often due to a microstrip antenna which is used for microwave excitation of
spin waves. In this context, the metalization plays the role of a mirror for the incoming light passing through the
sample (see Fig.~\ref{Bragg}). The wavevectors can be selected by varying the angle $\Theta_{B}$
between the sample and the probing beam, see Eq.~\eqref{eq1}. Theoretically, wavevectors up to $2.36 \cdot 10^{5}$~rad/cm
for $\Theta_{B\mathrm{max}} = 90^\circ$ are accessible using green probing light with a wavelength of 532~nm. Changing $\Theta_{B}$ is mostly realized by the rotation of the sample about the direction of the externally applied
magnetizing field. This works if the orientation of the magnetization is perpendicular to the direction of the
spin-wave wavevector. However, if the magnetization is parallel to the wavevector of the propagating spin
waves, the whole magnet and sample has to be tilted simultaneously to ensure the same conditions of magnetization.

Here we report on a setup capable of both rotating the sample inside the magnet and turning the whole magnet system
with respect to the probing beam at the same time. Due to geometrical constraints of the setup, $\Theta_{B\mathrm{max}}$ can be varied  up to $60^\circ$ and therefore dipole and exchange-dominated spin waves with wavenumbers up to
$\pm~2.04\cdot10^{5}$~rad/cm are accessible in both rotation directions. 


\section{Setup}

\begin{figure}
\begin{center}
\scalebox{1}{\includegraphics[width=8.5 cm, clip]{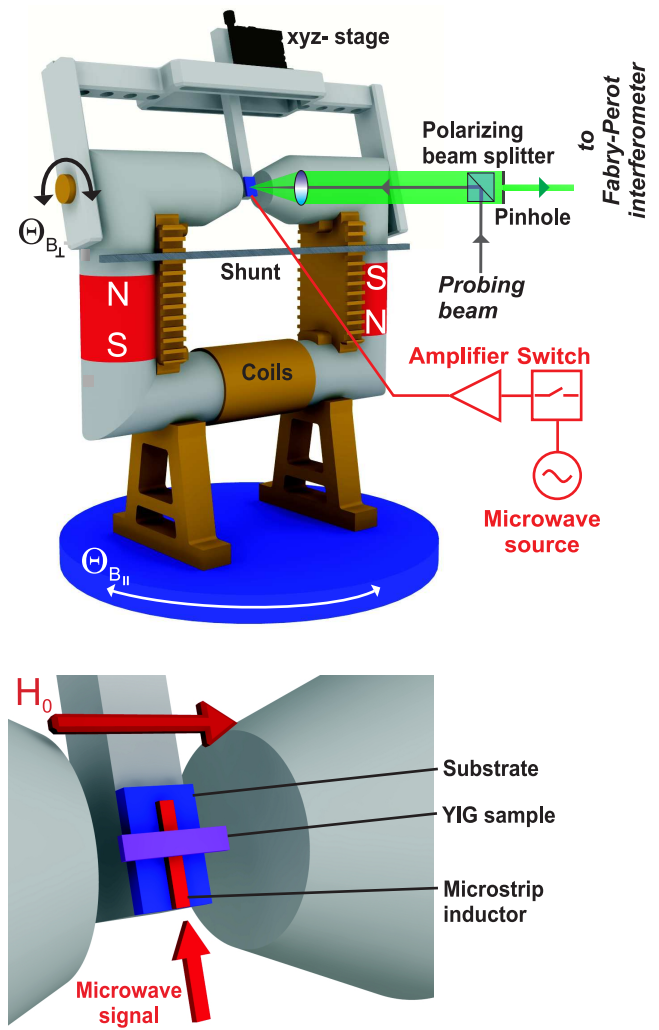}}
\end{center}
\caption{\label{Setup} (Color online) Experimental setup.
The main part of the setup consists of a yoke with two permanent magnets which is mounted on top of a rotating plate in order
to vary the angle $\Theta_{B\parallel}$. The sample itself is mounted on a freely pivotable lever with a small xyz-stage for precise adjustment.
Changing the angle $\Theta_{B\perp}$ will result in a selection of the in-plane components of wavevectors oriented perpendicular
to the bias field. The tuning of the field between the two poles is realized by iron shunts mounted parallel to the poles and a small electromagnet.
The sample is placed on a 50~$\mu$m wide microstrip antenna which is fabricated on top of an aluminum nitride substrate.
A microwave source, a switch, and a power amplifier are directly connected to the antenna in order to drive the spin-wave system.}
\end{figure}

Figure~\ref{Setup} shows the experimental setup and the enlarged area of the sample. The main part of the setup is a
yoke which is mounted on top of a rotating plate in order to vary the angle $\Theta_{B\parallel}$ between the sample
and the probing beam in the horizontal plane. By varying this angle, spin waves which are
oriented parallel to the external magnetic field ${\bf H}_{0}$ can be investigated. The sample itself is mounted on a freely pivotable lever
using a small xyz-stage, which allows precise positioning of the sample. The angle $\Theta_{B\perp}$ between the
lever and the probing light can be set with the same accuracy as $\Theta_{B\parallel}$, $0.1^\circ$ respectively. By
changing $\Theta_{B\perp}$ the in-plane components of wavevectors oriented perpendicular to the bias field
can be selected and investigated. The yoke itself consists of two NdFeB-permanent magnets and an iron alloy
(vacofer S1) in order to satisfy the demands of long-term stability of the bias magnetic field in combination with
a sufficiently large field strength of 2100~Oe. The poles of the yoke are cone-shaped so that the optical beam
path is not restrained. In addition, a brass frame has been
attached to the yoke in order to mount small iron bars (shunts) parallel to the poles. These iron bars partially
redirect the magnetic flux and effectively control the magnetic field in the pole gap. For the automated fine-tuning of the
bias magnetic field a small electric coil in combination with a Hall sensor have also been installed. Thus, the magnetic field strength can be tuned from 1300--2100~Oe, with an accuracy of 0.5~Oe. The probing light source is a single-mode solid
state laser with 532~nm wavelength. The light coming from the source is redirected by a polarizing beamsplitter cube
and sent through an objective to the sample. The benefit of using a polarizing beamsplitter is that the
polarization of inelastically scattered light is turned by $90^\circ$ with respect to the probing light\cite{Cottam}. Thus, the
inelastically scattered light can be separated from the probing light. Additionally a pinhole is placed in the beam of
the inelastically scattered light to assure that only the light coming directly through the middle of the objective is
passing to the interferometer.

\section{Experimental data}
\begin{figure}
\begin{center}
\scalebox{1}{\includegraphics[width=8 cm, clip]{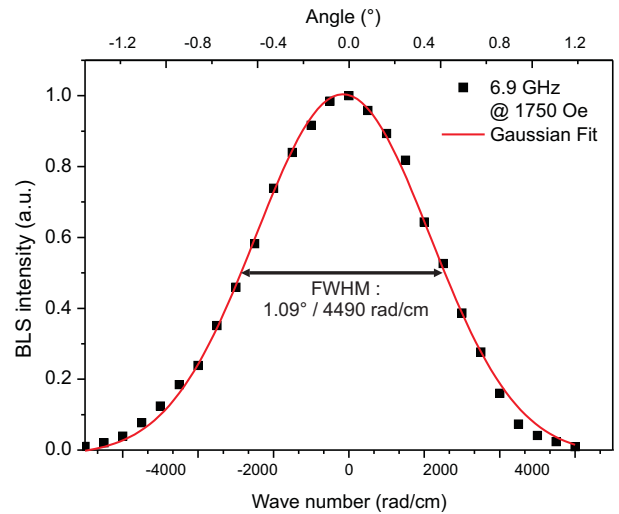}}
\end{center}
\caption{\label{FMR} (Color online) In order to determine the minimum resolution a dipolar spin wave is excited slightly below the FMR.
Its wavevector distribution shows a Gaussian profile with a
FWHM (Full Width Half Maximum) of $1.09^\circ$ which corresponds to a $\Delta{k} = 4490$~rad/cm.}
\end{figure}

The setup is designed for  wavevector-selective spectroscopy of thermally activated and externally excited
spin waves. Here, in order to present the functional capability of the setup, the probing of an externally driven magnon
gas in a YIG sample of 5~$\mu$m thickness placed on a 50~$\mu$m wide microstrip antenna is presented. The antenna
itself is fabricated on top of an aluminum nitride substrate which was used due to its excellent thermal conductivity to suppress
heating effects caused by high microwave power. A circuit consisting of a microwave source, a
switch, and a power amplifier is directly connected to the antenna. This microwave circuit enables  to
pump magnons in different parts of the spin-wave spectrum, i.e. both dipole-dominated and exchange-dominated spin waves.
The wavevector resolution of the setup depends on the errors in the determination
of the angles $\Theta_{B\parallel}$ and $\Theta_{B\perp}$.
Using Eq.~\eqref{eq1} the error condition
\begin{equation}\label{eq3}
\Delta{k} = 2k_{0}\cos(\Theta_{B})\Delta{\Theta_{B}}
\end{equation}
is obtained. Apparently, the error reaches its maximum value for small angles. The error $\Delta{k}$ for
$\Theta_{B} = 0^\circ$ is 4120~rad/cm, while for $\Theta_{B\mathrm{max}} = 60^\circ$ it is 206~rad/cm. Experimentally,
the minimum resolution of the setup can therefore be determined by measuring the wavevector distribution
for continuous spin waves excited near the ferromagnetic resonance (FMR), with a wavevector close to zero. Fig.~\ref{FMR} shows the wavevector
distribution for \textit{long-wavelength} dipolar dominated backward volume magnetostatic spin waves (BVMSW) with the
direction of propagation parallel to the bias magnetic field, so that $\Theta_{B\parallel}$ was varied. The measurement was performed at a magnetic field of 1750~Oe. The FMR frequency was independently experimentally determined to be 6.96~GHz by the observation of the microwave energy absorption using a network analyzer. The microwave source was operated in continuous-wave mode with an output power of 5~mW.  The frequency $f_{s}$ of the BVMSW is 6.90~GHz, slightly below the frequency of the FMR. It is clearly visible that the intensity as a function of angle $\Theta_{B\parallel}$ can be well fitted by a Gaussian profile with a FWHM (full width at half maximum) of $1.09^\circ$, which corresponds to a $\Delta{k}=4490$~rad/cm. For the setup this value is regarded as the minimum resolution.

\begin{figure}
\begin{center}
\scalebox{1}{\includegraphics[width=8.5 cm, clip]{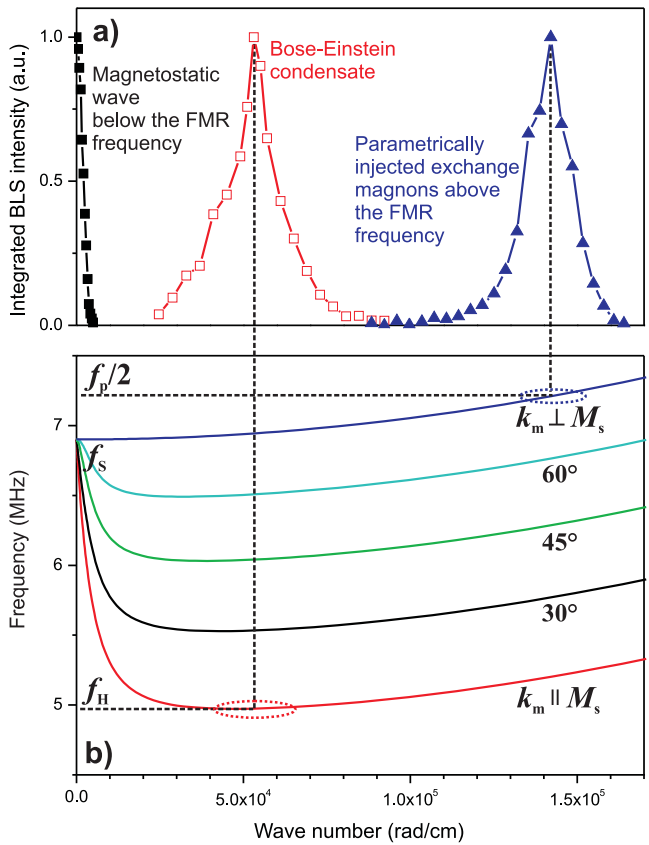}}
\end{center}
\caption{\label{Combined} (Color online) (a) Result of the measurements of different spectral areas: Magnetostatic waves below the FMR, directly
injected exchange magnons at $(f_\mathrm{p}/2, k_{\perp})=(7.2$~GHz, $1.42\cdot10^{5}$~rad/cm) and BEC of magnons at the lowest energy state $(f_H, k_{\|})=(5$~GHz, $5.2\cdot10^{4}$~rad/cm).
(b) Spin-wave dispersion relation and spectral positions of the excited magnons. An excellent agreement between measured and theoretically predicted spectral positions of both magnon groups is clearly visible here. Additionally, the dispersion branches for an angle of $30^\circ$, $45^\circ$, $60^\circ$ between the wavevector of the spin waves and the magnetization are shown.}
\end{figure}

To verify the \textit{short-wavelength} selectivity of the presented setup it is necessary to have
reference magnons which are narrow both in frequency and in wavevector space.
However, the short-wavelength magnons can not be excited by direct force excitation. The spatial variations of a
microwave magnetic field of the microstrip antenna are too smooth in comparison with desired variations of dynamic
magnetization. This is why we used instead the technique of parallel parametric microwave pumping \cite{Melkov} in order to create
these high wavevector magnons. In this process microwave photons split into frequency-degenerate pairs of phase-correlated
magnons at half of the pumping frequency with opposite wavevectors. In accordance with theoretical
predictions \cite{Lvov} competing interactions of the frequency-degenerate magnon pairs lead to the establishing of
only one magnon group with very well defined position in $\omega - k$ space. The orientation of the wavevector of
the magnons corresponding to this group is \textit{perpendicular} to the magnetization, i.e. $\Theta_{B\perp}$ has to
be varied to analyze the wavevector distribution.

At the same time, the parametric pumping process is not effective for injection of magnons propagating in a direction parallel to the magnetization \cite{Timo_APL}. However, the recent observation of the BEC of magnons in a parametrically
driven magnon gas \cite{Dem06} has shown that the magnons from the initially pumped group thermalize over the spin-wave
spectrum and condense at the lowest energy state. Thus, a strong spectral peak of inelastically scattered light is
observable at the bottom of the spin-wave frequency spectrum. As the spin waves at the bottom of the spectrum propagate
\textit{parallel} to the magnetization, $\Theta_{B\parallel}$ needs to be varied in this case.

As previously in the experiment with parametrically pumped magnons, the magnetizing field of 1750~Oe is applied in
perpendicular direction with respect to the microstrip. The pumping frequency is $f_\mathrm{p}=14.2$~GHz. The pumping power is
10~mW. Microwave pumping was applied in pulses of 2~$\mu$s time length with a repetition time of
20~$\mu$s. This is done in order to minimize the heating of the sample, which would result in a change of the saturation magnetization and the exchange stiffness
constant.

Figure~\ref{Combined}~(a) shows the result of $k$-vector resolved measurements of these two spectral areas: directly
injected exchange magnons at $(f_\mathrm{p}/2, k_{\perp})=(7.2$~GHz, $1.42\cdot10^{5}$~rad/cm) and BEC of magnons at the lowest energy state $(f_H, k_{\|})=(5$~GHz, $5.2\cdot10^{4}$~rad/cm). The spin-wave
dispersion relations, which we calculated using a recently developed microscopic approach \cite{Kreisel}, are presented in
Fig.~\ref{Combined}~(b). These dispersions were obtained by the numerical diagonalization of an effective spin Hamiltonian for YIG in linear spin wave theory. The excellent agreement between measured and theoretically predicted spectral positions of
both magnon groups is clearly visible here. At the same time one can mark the unexpectedly wide spreadings of magnon
wave numbers in both of these cases. These spreadings are significantly larger than the maximal measurement error (see
the wavevector distributions for long-wavelength dipolar dominated spin waves in Fig.~\ref{Combined}~(a) as well as
Fig.~\ref{FMR}) and must be understood as a result of the intrinsic processes in a magnon gas. Thus, we associate the
spreading of the parametrically pumped exchange magnons with inelastic 4-magnon scattering processes, which are
responsible for thermalization of a parametrically driven magnon gas \cite{Dem06}. The wide magnon spreading near the
BEC can be interpreted at this stage of the research as manifestation of a mixture of the magnon
condensate with the usual gaseous magnon states.

\section{Conclusion}

In conclusion, we have improved the wavevector resolving technique used in Brillouin light scattering spectroscopy of
magnetic media by two-dimensional access to the area of exchange dominated magnetic excitations. We have achieved this by
a combined experimental setup with a rotatable magnet enabling the wide-range wavevector resolution
\textit{along} the bias magnetic field and a pivotable sample holder which is placed between the magnetic poles which
permits the wavevector probing \textit{across} the bias field. The maximum error in measurement is determined by
analyzing the wavevector distributions of dipolar magnons excited near the frequency of ferromagnetic resonance in an
in-plane magnetized YIG film. The applicability of the developed technique to the probing of exchange areas of the
spin-wave spectrum was evidenced using a parametrically driven magnon gas in the same YIG sample.
Comparison with the established theory shows an excellent agreement between the calculated and experimentally measured
spectral positions of parametrically injected magnons and the BEC of magnons. The detailed
experimental and theoretical study of the discovered wide spreading of magnons from both groups in the reciprocal space 
is a subject of future research.

\begin{acknowledgments}
We thank F. Sauli for useful discussions and acknowledge the financial support
by the DFG within the SFB/TRR 49.
\end{acknowledgments}

\end{document}